\documentclass[sigconf]{acmart}
\acmConference[ESEC/FSE 2017]{Submitted to 11th Joint Meeting of the European Software Engineering Conference and the ACM SIGSOFT Symposium on the Foundations of Software Engineering}{4--8 September, 2017}{Paderborn, Germany}

\usepackage{booktabs} % For formal tables
\usepackage{algorithm, algorithmicx} % pseudo code
\usepackage{algpseudocode}% pseudo code

\usepackage{multirow}
\usepackage{tabu}
\usepackage{enumitem} % handle itemize, enumerate 
\usepackage{breakcites}
\usepackage{tabularx}
\usepackage{balance}
\usepackage{xcolor}
\usepackage[framemethod=tikz]{mdframed}
\usepackage{lipsum}
\usetikzlibrary{shadows}
\usepackage{mathtools}
\usepackage{graphics}
\newmdenv[
tikzsetting= {fill=white},
linewidth=1pt,
roundcorner=2pt, 
shadow=false
]{myshadowbox}

\usepackage{color,colortbl}
\usepackage{xcolor}
\definecolor{lightgray}{gray}{0.8}

\makeatletter
\let\th@plain\relax
\makeatother

\usepackage[framed]{ntheorem}
\usepackage{framed}
\usepackage{tikz}
\usetikzlibrary{shadows}

\definecolor{Gray}{rgb}{0.88,1,1}
\definecolor{Gray}{gray}{0.85}
\theoremclass{Lesson}
\theoremstyle{break}
\tikzstyle{thmbox} = [rectangle, rounded corners, draw=black,
 fill=Gray,  drop shadow={fill=black, opacity=1}]

\newcommand{\bi}{\begin{itemize}[leftmargin=0.4cm]}
\newcommand{\ei}{\end{itemize}}
\newcommand{\be}{\begin{enumerate}}
\newcommand{\ee}{\end{enumerate}}

\newcommand{\fig}[1]{Figure~\ref{fig:#1}}
\newcommand{\tab}[1]{Table~\ref{tab:#1}}

\newshadedtheorem{lesson}{Result}

\setcopyright{rightsretained}
\setcopyright{none}
\acmDOI{10.1145/3106237.3106257}
\acmISBN{978-1-4503-5105-8/17/09}
\acmConference[ESEC/FSE'17]{2017 11th Joint Meeting of the European Software Engineering Conference and the ACM SIGSOFT Symposium on the Foundations of Software Engineering}{September 4-8, 2017}{Paderborn, Germany} 
\acmYear{2017}
\copyrightyear{2017}
\acmPrice{15.00}

\begin{document}
\title{Revisiting Unsupervised Learning for Defect Prediction}
% \titlenote{Produces the permission block, and
% \subtitle{Extended Abstract}
% \subtitlenote{The full version of the author's guide is available as
%   \texttt{acmart.pdf} document}

\author{Wei Fu,  Tim Menzies}
\affiliation{%
  \institution{ Com. Sci., NC State, USA}
}
\email{wfu@ncsu.edu, tim.menzies@gmail.com}

% \author{Wei Fu}
% % \orcid{1234-5678-9012}
% \affiliation{%
%   \institution{Computer Science Department, North Carolina State University}
%   \streetaddress{890 Oval Drive}
%   \city{Raleigh} 
%   \state{North Carolina} 
%   \postcode{27606}
% }
% \email{wfu@ncsu.edu}

% \author{Tim Menzies}
% \affiliation{%
%   \institution{Computer Science Department, North Carolina State University}
%   \streetaddress{890 Oval Drive}
%   \city{Raleigh} 
%   \state{North Carolina} 
%   \postcode{27606}
% }
% \email{tim.menzies@gmail.com}

\begin{abstract}
Collecting quality  data from software projects
can be time-consuming and expensive. Hence,
	 some researchers explore ``unsupervised'' approaches
	 to quality  prediction that does not require labelled data.  An alternate technique is to use ``supervised'' approaches
	 that learn models from project data labelled with, say, 
	 ``defective'' or ``not-defective''.
	Most researchers use these supervised models
	since, it is argued, they can
	 exploit more 
	knowledge of the projects. 

	At FSE'16, Yang et al.  reported startling results
	where 
	unsupervised defect predictors 
	outperformed supervised
	predictors for effort-aware just-in-time defect prediction.
	If confirmed, these results  would
lead to a dramatic simplification of a seemingly complex task
(data mining) that is
widely explored in the software engineering  literature.
 
	This paper repeats and refutes those results as follows.
	(1)~There is much variability in the efficacy of the Yang et al.
	predictors so even with their approach, some supervised data is required
	to prune  weaker predictors away.
	(2)~Their findings were grouped across
	$N$  projects. When we repeat their analysis
	on a project-by-project basis, supervised 
	predictors are seen to work better.

Even though this paper rejects the specific conclusions of Yang et al.,
	we still endorse their  general goal. In our our experiments, supervised
	predictors did not perform outstandingly
	better than unsupervised ones for effort-aware just-in-time defect prediction.
	Hence, they may indeed be some
	combination of unsupervised
	learners to achieve comparable performance to supervised ones.
	We therefore encourage others to work in this promising area.
\end{abstract}



% We no longer use \terms command

\keywords{Data analytics for software engineering,
software repository mining,
empirical studies, defect prediction}

\maketitle

% \pagestyle{headings}
% \setcounter{page}{1}
%\pagenumbering{arabic} %XXX delete before submission
% \input{samplebody-conf}
\section{Introduction}

This paper repeats and refutes recent
results from Yang et al.~\cite{yang2016effort} published at FSE'16.
The task explored by Yang et al. was {\em effort-ware just-in-time (JIT) software defect predictors}.
JIT defect predictors are built on code change level and could be used to conduct defect prediction
right before developers commit the current change. They report an unsupervised software quality prediction method that
achieved
better results than standard supervised methods. 
We repeated their study since, if their results were confirmed,
this would imply that decades of research into defect prediction~\cite{lessmann2008benchmarking,hassan2009predicting,fu2016differential,fu2016tuning,graves2000predicting,menzies2007data,hall2012systematic,nagappan2005use,jiang2013personalized,kim2007predicting,kim2008classifying,jing2014dictionary,lee2011micro,moser2008comparative,nam2013transfer,wang2016automatically,yang2016effort,kamei2013large}
had needlessly complicated an inherently simple task.

The standard method for software defect prediction is learning from labelled data. In
this approach, the historical log of known defects is learned
by a data miner. Note that this approach requires waiting until  a
historical log of defects is available; i.e. until after the code has been used for a while.
Another  approach, explored by Yang et al., uses
general background knowledge to sort the code, then inspect the
code in that sorted order. In their study, they assumed that more defects
can be found faster by first looking over all the ``smaller'' modules
(an idea initially proposed by Koru et al.~\cite{koru2009investigation}).
After exploring various methods
of defining ``smaller'', they report their approach finds more defects, sooner, than  supervised methods. 
These results are  highly remarkable:
\bi
\item
This approach does not require access to labelled
data; i.e. it can be applied just as soon as the code is written.
\item
It is  extremely simple: no
data pre-processing, no data mining, just simple sorting.
\ei
Because of the remakrable nature of these results,
this paper takes a second look at the Yang et al. results.
% Yang et al.
% use 12 metrics to build
% 12 unsupervised predictors. Based on their evaluation results, they claimed that many unsupervised predictors perform better than Kamei et al.'s supervised predictor~\cite{yang2016effort}. However, since the unsupervised predictors
% are built on the testing data, to deploy unsupervised predictors on actual software
% project, we still have the following questions:
% \bi 
% \item Do all the $12$ unsupervised predictors perform equally well so that we can randomly pick one and apply to future projects?
% \item If not, without any prior knowledge on the future project,  which unsupervised predictor we have to use?
% \ei
% 
% According to our best knowledge, Yang et al. did not shed lights on these
% issues in ~\cite{yang2016effort}. Therefore, in this paper, we empirically
% study how to make such simple unsupervised predictors work on actual
% software projects. Specifically, we set the following research questions:
We ask three questions:

\textbf{RQ1: Do all unsupervised predictors perform better than supervised predictors?}

The reason we ask this question is that if the answer is ``yes", then we can simply select any unsupervised predictor built from the the change metrics as Yang et al suggested without using any supervised data; if the answer is ``no'', then we must apply some techniques to select best predictors and remove the worst ones. However, our results show that, when projects are explored separately, the majority of the unsupervised predictors learned by Yang et al. perform worse than supervised predictors. 

Results of RQ1 suggest that after  building multiple predictors using unsupervised methods,
it is required to prune the worst predictors and only better ones should be used for future prediction. However, with Yang et al. approach, there is no way to tell which unsupervised predictors will perform better without access to the labels of testing data. To test that speculation,
we built a new learner, {\it OneWay},  that  uses supervised training data to remove all but one of the Yang et al. predictors. Using this learner, we asked:

\textbf{RQ2: Is it beneficial to use supervised data to prune away all but one of the Yang et al. predictors?}

Our results showed that {\it OneWay} nearly always outperforms the unsupervised predictors found by Yang et al. The success of {\it OneWay} leads to one last question:

\textbf{RQ3: Does {\it OneWay} perform  better than more complex standard supervised learners?} 

	Such standard supervised learners include Random Forests, Linear Regression, J48 and IBk (these learners were selected based on prior results by ~\cite{menzies2007data,lessmann2008benchmarking,kamei2013large,hall2012systematic}).
We find that in terms 
of {\it Recall} and $P_{opt}$ (the metric preferred by Yang et al.), {\it OneWay} performed better than standard supervised predictors.  Yet measured in terms of
 {\it Precision}, there was no advantage to {\it OneWay}.

From the above, we make an opposite conclusion to Yang et al.; i.e.,
there are  clear advantages to use supervised approaches over unsupervised ones.
We explain the difference between our results and their results as follows:
\bi
\item Yang et al. reported averaged results across all projects;
\item We offer a more detailed analysis on a project-by-project basis.
\ei
The rest of this paper is organized as follows. Section~\ref{comment} is a commentary on Yang et al. study and the implication of this paper. Section ~\ref{background} describes the background and related work on defect prediction.
Section~\ref{method} explains the effort-ware JIT defect prediction methods investigated in this study. Section ~\ref{experiment} describes the
experimental settings of our study, including research questions that motivate our study, data sets and experimental design.
Section~\ref{results} presents the results. Section ~\ref{threats} discusses the threats to the validity of our study.  Section ~\ref{conclusion} presents the conclusion and future work.

Note one terminological convention: in the following, we treat ``predictors'' and ``learners'' as synonyms.

\section{SCIENCE IN the 21{\lowercase{st}} Cenury}\label{comment}

While this paper is specific about effort-aware JIT defect prediction and the Yang et al. result, at another level this paper is also about science in the 21st century. 
 
In 2017, the software analytics community now has the tools, data sets, experience to explore a {\em bold } wider range of options. There are practical problems in exploring all those possibilities specifically, too many options. For example, in section 2.5 of \cite{kocaguneli2012exploiting}, Kocaguneli et al. list  12,000+ different ways of estimation by analogy. We have had some recent successes with exploring this space of options \cite{fu2016tuning} but only after the total space of options is reduced by some {\em initial} study to a manageable set of possibilities. Hence, what is needed are {\em initial} studies to rule our methods that are generally unpromising (e.g. this paper) before we apply second level hyper-parameter optimization study that takes the reduced set of options.
 
Another aspect of 21st century science that is highlighted by this paper is the nature of repeatability. While this paper disagrees the conclusions of Yang et al., it is important to stress that their paper is an excellent example of good science that should be emulated in future work. 

Firstly, they tried something new. There are many papers
in the SE literature about defect prediction. However, compared to most
of those, the Yang et al. paper is bold and stunningly original.

Secondly, they made all their work freely available. Using
the ``R'' code they placed online, we could reproduce their result,
including all their graphical output, in a matter of days.  
Further, using that code as a starting point, we could rapidly
conduct the extensive experimentation that leads to this paper. 
This is an excellent example of the value of open science.

Thirdly, while we assert their answers were wrong, the question
they asked is important and should be treated as an open and urgent
issue by the software analytics  community. 
In our experiments, supervised
	predictors performed better than unsupervised--
	but not   outstandingly
	better  than unsupervised. 
	Hence, they may indeed be some
	combination of unsupervised
	learners to achieve comparable performance to supervised.
	Therefore, even though we reject the specific conclusions of Yang et al.,
	we still endorse the question they asked  strongly
	and encourage others to work in this area.

\section{Background and Related Work}\label{background}

\subsection{Defect Prediction}

As soon as people started programming, it became apparent
that programming was an inherently buggy process. As recalled
by Maurice Wilkes~\cite{wilkes1985memoirs}, speaking of his programming experiences from the early 1950s: ``It was on one of my journeys between the EDSAC room and the punching equipment that `hesitating at the angles of stairs' the realization came over me with full force that a good part of the remainder of my life was going to be spent in finding errors in my own programs.''

It took  decades to gather the experience required to quantify the size/defect relationship. In 1971, Fumio Akiyama~\cite{akiyama1971example} described the first known ``size'' law, saying the number of defects
D was a function of the number of LOC where \mbox{$D=4.86+0.018*\mathit{LOC}$}.

In 1976, Thomas McCabe argued that the number of LOC was less important than the complexity of that code~\cite{mccabe1976complexity}. He argued
that code is more likely to be defective when his ``cyclomatic complexity'' measure was over 10. Later work used data miners to build defect predictors
that proposed thresholds on multiple measures~\cite{menzies2007data}. 

Subsequent research showed 
that software bugs are not distributed evenly across
a system. Rather, they seem to clump in small corners of the code. For example,
Hamill et al.~\cite{hamill09} report studies with (a)~the GNU C++ compiler
where half of the files were never implicated in issue reports while 10\% of
the files were mentioned in half of the issues. Also,
Ostrand et al.~\cite{Ostrand:2004} studied
(b)~AT\&T data and reported that 80\% of the bugs reside in 20\% of the files.
Similar ``80-20'' results have been observed in (c)~NASA systems~\cite{hamill09} as well
as (d)~open-source software~\cite{koru2009investigation} and (e)~software
from Turkey~\cite{misirli2011ai}. 

Given this skewed distribution of bugs,  a
cost-effective   quality assurance approach is to  sample
across a software system, then focus  on regions reporting some
bugs.
Software defect predictors built from data miners are 
 one way to implement such a sampling policy. While their conclusions
are never 100\% correct, they can be used to suggest where to focus more expensive  methods such as elaborate manual
review of source code~\cite{Shull:2001}; symbolic execution checking~\cite{paaareanu2008combining}, etc.
For example, Misirli et al.~\cite{misirli2011ai} report studies where the guidance offered by
defect predictors:
\bi
\item
Reduced the effort required for
software inspections in some Turkish software companies by
72\%;
\item
While, at the same time, still being able to find the 25\% of the files that
contain 88\% of the defects.
\ei
Not only do static code defect predictors perform well compared to manual methods,
they also are competitive with certain automatic methods.
A recent study at ICSE'14, Rahman et al.~\cite{rahman2014comparing} compared
(a) static code analysis tools FindBugs, Jlint, and Pmd and (b)
static code defect predictors
(which they called ``statistical defect prediction'') built using logistic regression.
They found  no significant differences in the cost-effectiveness
of these  approaches. Given this equivalence, it is significant to note that 
static code defect prediction can be quickly adapted to new languages by building lightweight
parsers that extract static code metrics. The same is not true for   static code analyzers-- these need  extensive modification before they can be used on new
languages.

% Defect prediction for software projects has been well studied 
% over the past decade and lots of researchers focus on proposing new techniques to improve predictors' performance
% or make predictors more feasible for practical situation~\cite{lessmann2008benchmarking,hassan2009predicting,fu2016tuning,graves2000predicting,menzies2007data,hall2012systematic,nagappan2005use,jiang2013personalized,kim2007predicting,lee2011micro,moser2008comparative,wang2016automatically}. 
To build such defect predictors,
  we measure the complexity of
software projects using McCabe metrics, Halstead's effort metrics and  CK object-oriented code mertics~\cite{kafura1987use,chidamber1994metrics,mccabe1976complexity,halstead1977elements} at
a coarse granularity, like file or package level. With the collected data instances along with the corresponding labels~(defective or non-defective), we can  build defect prediction models using supervised learners such as Decision Tree, Random Forests, SVM, Naive Bayes and Logistic Regression~\cite{khoshgoftaar2001modeling,khoshgoftaar2003software,khoshgoftaar2000balancing,menzies2007data,lessmann2008benchmarking,hall2012systematic}. After that, such trained defect predictor can be applied
to predict the defects of future projects. 
\subsection{Just-In-Time Defect Prediction}
Traditional defect prediction has some drawbacks such as prediction at a coarse granularity and started at very late stage of software development circle~\cite{kamei2013large}, whereas in JIT defect prediction paradigm, the defect predictors are built on code change level, which could easily help developers  narrow down the code for inspection and JIT defect prediction could be conducted right before developers commit the current change. JIT defect prediction becomes a more practical method for practitioners to carry out.

Mockus et al.~\cite{mockus2000predicting} conducted the first study to predict 
software failures on a telecommunication software project, 5ESS,
by using logistic regression on 
data sets consisted of change metrics of the project. Kim et al. ~\cite{kim2008classifying}
further evaluated the effectiveness of change metrics on open source projects. In their study, they proposed to apply support vector machine to build a defect predictor based on software change metrics, where on average they achieved $78\%$ accuracy and $60\%$ recall. Since training data might
not be available when building the defect predictor, Fukushima et al.~\cite{fukushima2014empirical} introduced cross-project paradigm into JIT defect prediction. Their results showed that using data from other projects to build JIT defect predictor is feasible. 

 Most of the research into defect prediction   does not consider the effort\footnote{Effort means time/labor required to inspect total number
of files/code predicted as defective.} required to inspect the code predicted to be defective. Exceptions to this
 rule include the work of Arishom and Briand~\cite{Arisholm:2006},
 Koru et al.~\cite{koru2009investigation} and Kamei et al.~\cite{kamei2013large}.
 Kamei et al.~\cite{kamei2013large} conducted a large-scale study on the effectiveness of JIT defect prediction, where they claimed that using $
20\%$ of efforts required to inspect all changes, their modified linear regression model~(EALR) could
detect $35\%$ defect-introducing changes. Inspired by Menzies et al.'s ManualUp model~(i.e., small size of modules inspected first)~\cite{menzies2010defect}, Yang et al.~\cite{yang2016effort} proposed to build $12$ unsupervised defect predictors by sorting the reciprocal values of $12$ different change metrics on each testing data set in descending order. They
reported that with $20\%$ efforts, many
unsupervised predictors perform better than state-of-the-art supervised predictors.

\section{Method}\label{method}

\subsection{Unsupervised Predictors} \label{unsupervised}
In this section, we describe the effort-aware 
just-in-time unsupervised defect predictors proposed by Yang et al. ~\cite{yang2016effort},
which serves as a baseline method in this study. As described by Yang et al.~\cite{yang2016effort},  their simple unsupervised
 defect predictor is built on change metrics as shown in \tab{CM}. These
 14 different change metrics can be divided into 5 dimensions~\cite{kamei2013large}:
 
 \bi
 \item Diffusion: NS, ND, NF and Entropy.
 \item Size: LA, LD and LT.
 \item Purpose: FIX.
 \item History: NDEV, AGE and NUC.
 \item Experience: EXP, REXP and SEXP.
 \ei

%%%%%%% ================table--begin======================
\begin{table}[!htp]
\caption{Change metrics used in our data sets.}\label{tab:CM}
\centering
\resizebox{0.48\textwidth}{!}{
\begin{tabular}{l|l}
\hline
\rowcolor{lightgray}
Metric & Description  \\\hline
NS & Number of modified subsystems~\cite{mockus2000predicting}.\\
ND & Number of modified directories~\cite{mockus2000predicting}. \\
NF & Number of modified files~\cite{nagappan2006mining}. \\
Entropy & Distribution of the modified code across each file~\cite{d2010extensive,hassan2009predicting}.\\\hline
LA & Lines of code added~\cite{nagappan2005use}.\\
LD & Lines of code deleted~\cite{nagappan2005use}.\\
LT & Lines of code in a file before the current change~\cite{koru2009investigation}.\\\hline
FIX & Whether or not the change is a defect fix~\cite{guo2010characterizing,yin2011fixes}. \\\hline
NDEV & Number of developers that changed the modified files~\cite{matsumoto2010analysis}. \\
AGE & The average time interval between the last and the current \\
    & change~\cite{graves2000predicting}. \\
NUC & The number of unique changes to the modified files~\cite{d2010extensive,hassan2009predicting}. \\ \hline
EXP & The developer experience in terms of number of changes~\cite{mockus2000predicting}.\\
REXP & Recent developer experience~\cite{mockus2000predicting}. \\
SEXP & Developer experience on a subsystem~\cite{mockus2000predicting}. \\\hline
\end{tabular}}
\end{table}
%%%%%%% ================table--end======================

The {\it diffusion} dimension characterizes how a change is 
distributed at different levels of granularity.
As discussed by Kamei et al.~\cite{kamei2013large},
a highly distributed change is harder to keep track
and more likely to introduce defects. The {\it size} dimension
characterizes the size of a change and it is believed that
the software size is related to defect proneness
~\cite{nagappan2005use,koru2009investigation}. Yin et al.~\cite{yin2011fixes}
report that the bug-fixing process can also introduce new bugs. 
Therefore, the {\it Fix} metric could be used as a defect evaluation metric.
The {\it History} dimension includes some historical information
about the change, which has been proven to be a good defect indicator ~\cite{matsumoto2010analysis}.
For example, Matsumoto et al. ~\cite{matsumoto2010analysis} find that the files
previously touched by many developers are likely to contain more defects.
The {\it Experience} dimension describes the experience of software programmers
for the current change because Mockus et al.~\cite{mockus2000predicting} show that
more experienced developers are less likely to introduce a defect. More details
about these metrics can be found in Kamei et al's study~\cite{kamei2013large}.

In Yang et al.'s study, for each change metric $M$ of testing data, 
they build an unsupervised predictor that ranks all the changes
based on the corresponding value of $\frac{1}{M(c)}$ in descending order, 
where $M(c)$ is the value of the selected change metric for each change $c$.
Therefore, the changes with smaller change metric values will ranked higher.
In all,  for each project, Yang et al. define  12 simple
unsupervised predictors~(LA and LD are excluded as Yang et al.~\cite{yang2016effort}).

%  The literature show that these factors perform well to in defect prediction
% \cite{mockus2000predicting,nagappan2006mining,nagappan2005use,
% koru2009investigation, hassan2009predicting,guo2010characterizing,
% matsumoto2010analysis,graves2000predicting,d2010extensive,kamei2013large}. 
% In particular, Kamei et al.\cite{kamei2013large}

\subsection{Supervised Predictors}
To further evaluate the unsupervised predictor, we selected some supervised predictors
that already used in Yang et al.'s work. 

As reported in both Yang et al.'s~\cite{yang2016effort}
and Kamei et al.'s~\cite{kamei2013large} work, EALR outperforms
all other supervised predictors for effort-aware JIT defect prediction.
EALR is a modified linear regression model~\cite{kamei2013large} and it predicts $\mathit{\frac{Y(x)}{Effort(x)}}$ instead of predicting $Y(x)$,
 where $Y(x)$ indicates whether this change
is a defect or not~($1$ or $0$) and $\mathit{Effort(x)}$ represents the effort required 
to inspect this change. Note that this is the same method to 
build EALR as Kamei et al. ~\cite{kamei2013large}.

In defect prediction literature, IBk~(KNN), J48 and Random Forests methods are simple yet widely used  as defect learners and have been proven to perform, if not best, quite
well for defect prediction~\cite{lessmann2008benchmarking,menzies2007data,hall2012systematic,turhan2009relative}. These three
learners are also used in Yang et al's study. For these
supervised predictors, $Y(x)$ was used as the dependant variable. 
For KNN method, we set $K=8$ according to Yang et al.~\cite{yang2016effort}.

\subsection{OneWay Learner}\label{OneWay}
% \wei{The following text probabily need to move to motivation part}

% As described in section \ref{unsupervised}, given a new project,
% Yang et al.'s unsupervised method builds 12
% learners for each project according to each predictor(change metric).
% However, without using some validation data and ground truth labels to test,
% we do not know which simple unsupervised predictor works well for this project.
% Generally, there are two situations need to consider, where

% \bi
% \item all 12 learners  perform better than supervised predictors.
% \item or only some learners  perform better than supervised predictors.
% \ei

% In the first situation, we have the confidence that there is no need to
% care about choosing which change metric to build unsupervised predictors on.
% Simply choosing one out of those 12 metrics and building a simple
% learner on it will work quite well for JIT effort-aware defect prediction. 
% However, in the second situation, we have to come up a way to
% figure out how to select good predictors. Otherwise, randomly selecting
% one feature to build an unsupervised predictor will not guarantee the performance.
% In other words, without any training data or supervised data, we can not even
% tell which situation fits our new project.

Based on our preliminary experiment results
shown in the following section, for the six projects
investigated by Yang et al, some of 12 unsupervised
predictors do perform worse than supervised predictors and there is 
no one predictor constantly working best on all project data.
This means we can not simply say which unsupervised predictor 
works for the new project before predicting on the testing data. In this case,
we need a technique to select the proper metrics to build defect predictors.

We propose {\it OneWay} learner, which is a supervised predictor built on
the implication of Yang et al's simple unsupervised predictors. 
%Since not all unsupervised predictors
% work equally well on all project and the missing step in Yang et al is to select the best predictor on a project-by-project basis, we propose to utilize the training data to select the best simple predictor.
The pseudocode for {\it OneWay}
is shown in Algorithm~\ref{alg:OneWay}. In the following description,
we use the superscript numbers to denote the line number in pseudocode. 

The general idea of {\it OneWay} is to use supervised training data to 
remove all but one of the Yang et al. predictors and then apply this trained
learner on the testing data. Specifically, {\it OneWay} firstly builds simple
unsupervised predictors from each metric on training data$^{L4}$, then
evaluates each of those learners in terms of evaluation metrics$^{L5}$,
like $P_{opt}$, {\it Recall}, {\it Precision} and {\it F1}. After that, if the desirable
evaluation goal is set, the metric which performs best on the corresponding
evaluation goal is returned as the best metric; otherwise, the metric
which gets the highest mean score over all evaluation metrics is returned$^{L9}$(In this study,
we use the latter one).
Finally, a simple predictor  is built only
on such best metric$^{L10}$ with the help of training data. Therefore, {\it OneWay} builds only one supervised predictor for each project using the local data instead of
12 predictors directly on testing data as Yang et al \cite{yang2016effort} .

\begin{algorithm}[!htp]
\small
\caption{Pseudocode for OneWay}
\begin{algorithmic}[1]
\Require $\mathit{data\_train}$, $\mathit{data\_test}$, $\mathit{eval\_goal} \in \{\mathit{F1, P_{opt}, Recall, Precision,...}\}$
\Ensure $\mathit{result}$
\vspace{2mm}
\Function{OneWay}{$\mathit{data\_train}$, $\mathit{data\_test}$, $\mathit{eval\_goal}$}
  \State $\mathit{all\_scores} \gets$ NULL
  \For{$\mathit{metric}$ in $\mathit{data\_train}$}
      \State $\mathit{learner}\gets$ buildUnsupervisedLearner($\mathit{data\_train}$, $\mathit{metric}$)
      \State $\mathit{scores} \gets $evaluate($\mathit{learner}$)\\
      \quad \qquad //$\mathit{scores}$ include all evaluation goals, e.g., $\mathit{Popt,F1,...}$
      \State $\mathit{all\_scores}$.append($\mathit{scores}$)
  \EndFor
  \State $\mathit{best\_metric} \gets$ pruneFeature($\mathit{all\_scores}$, $\mathit{eval\_goal}$)
  \State $\mathit{result} \gets$ buildUnsupervisedLearner($\mathit{data\_test}$, $\mathit{best\_metric}$)
  \State \Return $\mathit{result}$
\EndFunction
\Function{pruneFeature}{$\mathit{all\_scores}$, $\mathit{eval\_goal}$}
  \If{$\mathit{eval\_goal}$ == NULL}
    \State $\mathit{ mean\_scores} \gets$ getMeanScoresForEachMetric($\mathit{all\_scores}$)
    \State $ \mathit{best\_metric} \gets$ getMetric(max($\mathit{mean\_scores}$)) 
    \State \Return $\mathit{best\_metric}$
   \Else
     \State $\mathit{best\_metric} \gets$ getMetric(max($\mathit{all\_scores[``eval\_goal"]}$))
     \State\Return $\mathit{best\_metric}$
  \EndIf
\EndFunction
\end{algorithmic}
\label{alg:OneWay}
\end{algorithm}

\section{Experimental Settings}\label{experiment}

\subsection{Research Questions}
Using the above methods, we explore three questions:
\bi 
\item Do all unsupervised predictors perform better than supervised predictors?
\item Is it beneficial to use supervised data to prune all but one of the Yang et al. unsupervised predictors?
\item Does {\it OneWay} perform  better than more complex standard supervised predictors?
\ei

When reading the results from Yang et al.~\cite{yang2016effort},
we find that they aggregate performance scores of each learner on six projects, which might miss some information about how learners perform on each project. Are these unsupervised predictors working consistently across all the project data? If not, how would it look like? Therefore, in RQ1, we report results for each project separately.

Another observation is that even though Yang et al.~\cite{yang2016effort} propose that 
simple unsupervised predictors could work better than supervised predictors for effort-aware JIT defect prediction, one missing aspect of their report is how to select the most promising metric  to build a defect predictor. This is not an issue when all unsupervised predictors perform well but, as we shall see, this is not the case. As demonstrated below, given $M$ unsupervised predictors, only a small subset can be recommended. Therefore it is vital to have some mechanism by which we can down select from $M$ models to the $L \ll M$ that are 
useful. Based on this fact, we propose a new method, {\it OneWay}, which is the missing link in Yang et al.'s study~\cite{yang2016effort} and the missing final step they do not explore. Therefore,
in RQ2 and RQ3, we want to evaluate how well our proposed {\it OneWay} method performs compared to the unsupervised predictors and supervised predictors.

Considering our goals and questions, we reproduce Yang et al's results
and report for each project to answer RQ1. For RQ2 and RQ3, we implement our {\it OneWay}
method, and compare it with unsupervised predictors and supervised predictors on different projects in terms of various evaluation metrics.

\begin{table}[!htp]
    \centering
    \caption{Statistics of the studied data sets}
    \resizebox{0.48\textwidth}{!}{
    \begin{tabular}{@{}c|c c c c c@{}}
    \hline
        \multirow{3}{*}{Project} &
        \multirow{3}{*}{Period} & 
        \multirow{3}{*}{\begin{tabular}[c]{@{}c@{}}Total\\ Change\end{tabular}} & 
        \multirow{3}{*}{\begin{tabular}[c]{@{}c@{}}$\%$ of \\ Defects \end{tabular}} &
        \multirow{3}{*}{\begin{tabular}[c]{@{}c@{}}Avg LOC \\ per \\ Change\end{tabular}}  &
        \multirow{3}{*}{\begin{tabular}[c]{@{}c@{}} \# Modified \\ Files per\\ Change\end{tabular}} \\ \\  \\\hline
    Bugzilla & 08/1998 - 12/2006 & 4620 & 36\% & 37.5 & 2.3 \\
    Platform & 05/2001 - 12/2007 & 64250 & 14\%& 72.2 & 4.3 \\
    Mozilla & 01/2000 - 12/2006 & 98275 & 5\% & 106.5 & 5.3 \\
    JDT & 05/2001 - 12/2007 & 35386 & 14\% & 71.4 & 4.3 \\
    Columba & 11/2002 - 07/2006 & 4455 & 31\% & 149.4 & 6.2 \\
    PostgreSQL & 07/1996 - 05/2010 & 20431 & 25\% & 101.3 & 4.5 \\ \hline
    \end{tabular}}
    \label{tab:datasets}
\end{table}

\subsection{Data Sets}
In this study, we conduct our experiment using the same data sets as Yang et al.\cite{yang2016effort},
which are six well-known open source projects, Bugzilla, 
Columba, Eclipse JDT, Eclipse Platform, Mozilla and PostgreSQL.
These data sets are shared by Kamei et al.~\cite{kamei2013large}.
The statistics of the data sets are listed in ~\tab{datasets}.
From ~\tab{datasets}, we know that all these six data sets
cover at least 4 years historical information, and the longest one is PostgreSQL, which includes 15 years of data. The total changes for these six data sets
are from $4450$ to $98275$, which are sufficient for us to conduct
an empirical study. In this study, if a change introduces one or more defects then
this change is considered as defect-introducing change. The percentage of 
defect-introducing changes ranges from 5\% to 36\%. All the data and code used in this paper is available online\footnote{\url{https://github.com/WeiFoo/RevisitUnsupervised}}.

\subsection{Experimental Design}\label{exp_design}

The following principle guides the design of these experiments:\\
% \bi
% \item
 {\it Whenever there is a choice between methods, data, etc., we will always prefer the techniques used in Yang et al.~\cite{yang2016effort}.}\\
%  \ei
 By applying this principle, we can ensure that our experimental setup is the same as Yang et al.~\cite{yang2016effort}. This will increase the validity of our comparisons with that prior work.

When applying data mining algorithms to build predictive models,
one important principle is not to test on the data used in training. 
To avoid that, we used time-wise-cross-validation method which is also used by Yang et al.~\cite{yang2016effort}.
The important aspect of the following experiment is that
it ensures   that all testing data was  
created after training data.
Firstly, we sort all the changes in each project based on the commit date. 
Then all the changes that were submitted in the same month are grouped together.
For a given project data set that covers totally $N$ months history, when building
a defect predictor, consider a sliding window size of $6$,
\bi 
\item The first two consecutive months data in the sliding window,  $i$th and $i+1$th, are used as the training data to build supervised predictors and {\it OneWay} learner.
\item The last two months data in the sliding window, $i+4$th and $i+5$th, which are
two months later than the training data, are used as
the testing data to test the supervised predictors, {\it OneWay} learner and unsupervised predictors.
\ei
After one experiment, the window slides by ``one month'' data. By using this method, each training and testing data set
has two months data, which will include sufficient positive and negative instances
for the supervised predictors to learn. For any project that includes $N$ months data,
we can perform $N-5$ different experiments to evaluate our learners when $N$ is greater than 5.
For all the unsupervised predictors, only the testing data 
is used to build the model and evaluate the performance.

To statistically compare the differences between {\it OneWay} 
with supervised and unsupervised predictors,
we use 
% Scott-Knott procedure recommended by ~\cite{mittas2013ranking}.
% Scott-Knott first looks for a break in the sequence that maximizes the expected
%     values in the difference in the means before and after the break.
%     More specifically,  it  splits $l$ values into sub-lists $m,n$ in order to maximize the expected value of differences  in the observed performances before and after divisions. E.g. for lists $l,m,n$ of size $ls,ms,ns$ where $l=m\cup n$, Scott-Knott divides the sequence at the break that maximizes:
%      \[E(\Delta)=\frac{ms}{ls}abs(m.\mu - l.\mu)^2 + \frac{ns}{ls}abs(n.\mu - l.\mu)^2\]
% Scott-Knott then applies some statistical hypothesis test $H$ to check if $m,n$ are significantly different.
% If so, Scott-Knott then recurses on each division; 
% Otherwise, Scott-Knott terminates and returns the ranks
% of sub-lists.
Wilcoxon single ranked test to compare
the performance scores of the learners in this study the same as Yang et al.~\cite{yang2016effort}.
To control the false discover rate, the Benjamini-Hochberg~(BH) adjusted p-value 
is used to test whether two distributions are statistically significant
at the level of $0.05$~\cite{benjamini1995controlling, yang2016effort}. 
To measure the effect size of performance scores among {\it OneWay} and supervised/unsupervised predictors,
we compute Cliff's $\delta$ that is a non-parametric effect size measure~\cite{romano2006exploring}.
As Romano et al. suggested, we evaluate the magnitude of the effect size as follows:
negligible ($|\delta|<0.147$ ), small ($ 0.147\leq|\delta|<0.33$), medium ($0.33\leq|\delta|<0.474$ ), and large (0.474 $\leq|\delta|$)~\cite{romano2006exploring}.

\subsection{Evaluation Measures}

For effort-aware JIT defect prediction, in addition to evaluate how learners correctly predict a defect-introducing change, we have to take account  the efforts that are required to inspect prediction. Ostrand et al.~\cite{ostrand2005predicting}
report that given a project, $20\%$ of the files contain on average $80\%$  of all defects in the project.
Although there is nothing magical about the number $20\%$, it has been used as a cutoff
value to set the efforts required for the defect inspection when evaluating the defect learners ~\cite{yang2016effort,kamei2013large,mende2010effort,monden2013assessing}.
That is, given $20\%$ effort, 
how many defects  can be detected by the learner.
To be consistent with Yang et al, in this study, we restrict our efforts to $20\%$ of total efforts.

To evaluate the performance of effort-aware JIT defect
prediction learners in our study, 
we used the following 4 metrics: {\it Precision, Recall, F1} and $P_{opt}$,
which are widely used in defect prediction literature~\cite{menzies2007data,menzies2010defect,zimmermann2007predicting, kamei2013large,yang2016effort,monden2013assessing}.

% \vskip -2ex
\begin{eqnarray}\nonumber
Precision &=& \frac{True \, Positive}{True \, Positive +False \, Positive}\\ \nonumber
Recall &=& \frac{True \, Positive}{True \, Positive +False \, Negative}\\ \nonumber
F1 &=& \frac{2*Precision*Recall}{Recall + Precision}
\end{eqnarray}
where {\it Precision} denotes the percentage of actual defective changes to all the predicted changes and {\it Recall} is the percentage of predicted defective changes to all actual defective changes. {\it F1 } is a measure that combines both {\it Precision} and {\it Recall} which is the harmonic mean of {\it Precision} and {\it Recall}. 

% \vskip -2ex
\begin{figure}[!tph]
    \centering
    \includegraphics[width=.33\textwidth]{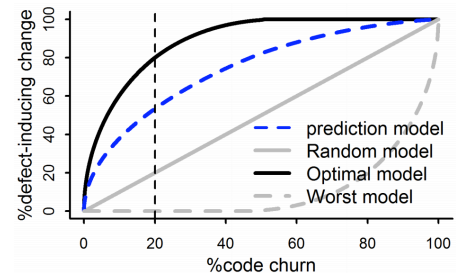}
    \caption{Example of an effort-based cumulative lift chart ~\cite{yang2016effort}.}
    \label{fig:delta}
\end{figure}

% \vskip -2ex

The last evaluation metric used in this study is $P_{opt}$,
which is defined as $1- \Delta_{opt}$, where $\Delta_{opt}$ is the area
between the effort~(code-churn-based) cumulative lift charts of the optimal
model and the prediction model (as shown in \fig{delta}). 
In this chart, the x-axis is considered as the percentage of required effort to inspect the change and the y-axis
is the percentage of defect-introducing change found in the selected change. In the optimal model, all the changes are sorted by the actual defect density in descending order, while for the predicted model, all the changes are sorted by 
the actual predicted value in  descending order.

According to Kamei et al. and Xu et al. ~\cite{yang2016effort,kamei2013large,monden2013assessing}, $P_{opt}$ can be normalized as follows:
\begin{eqnarray} \nonumber
P_{opt}(m) = 1- \frac{S(optimal)-S(m)}{S(optimal)-S(worst)}
\end{eqnarray}
where $S(optimal)$, $S(m)$ and $S(worst)$ represent the area of curve under the optimal model, predicted model, and worst model, respectively. Note that the worst model is built by sorting all the changes according to the actual defect density in ascending order. For any learner, it performs better than random predictor only if the ${P_{opt}}$ is greater than 0.5.

Note that, following the practices of Yang et al.~\cite{yang2016effort},
we measure {\it Precision}, {\it Recall}, {\it F1} and $P_{opt}$ at the effort = $20\%$ point. In this study, in addition to $P_{opt}$ and $ACC$ (i.e., {\it Recall}) that is used in Yang et al's work~\cite{yang2016effort}, we include 
{\it Precision} and {\it F1} measures and they provide more insights about all the learners evaluated in the study from very different perspectives, which will be shown in the next section.

\section{Empirical Results}~\label{results}
In this section, we present the experimental results to investigate how
simple unsupervised predictors work in practice and evaluate the performance
of the proposed method, {\it OneWay}, compared with supervised and unsupervised predictors.

Before we start off, we need a sanity check to see if we can fully 
reproduce Yang et al.'s results. Yang et al.~\cite{yang2016effort}
provide the median values of $P_{opt}$ and {\it Recall} for the EALR
model and the best two unsupervised models, LT and AGE, from the
time-wise cross evaluation  experiment. Therefore, we use those numbers to check
our results.

As shown in \tab{comp_Popt} and \tab{comp_Recall}, for unsupervised predictors,
LT and AGE, we get the exact same performance scores on all projects in terms
of {\it Recall} and ${\it P_{opt}}$. This is reasonable because unsupervised predictors
are very straightforward and easy to implement. For the supervised predictor, EALR, 
these two implementations do not have differences in ${\it P_{opt}}$, while the maximum
difference in  {\it Recall} is only ${0.02}$. Since the differences are quite small,
then we believe that our implementation reflects the details about EALR and unsupervised
learners in Yang et al.~\cite{yang2016effort}. 

For other supervised predictors used in this study, like J48, IBk, and  Random Forests,
we use the same algorithms from Weka package~\cite{hall2009weka} and set the same parameters as  used in Yang et al.~\cite{yang2016effort}.

\begin{table}[!htp]
    \centering
    \caption{ Comparison in $P_{opt}$: Yang's method~(A) vs. our implementation~(B)}
    % \resizebox{0.48\textwidth}{!}{
    \begin{tabular}{l|c c| c c| c c}

    \hline
        \multirow{2}{*}{Project} &
        \multicolumn{2}{c|}{EALR} &
        \multicolumn{2}{c|}{LT} &\multicolumn{2}{c}{AGE} \\
        \cline{2-7}
        &A & B& A&B&A&B \\
        \hline
        Bugzilla & 0.59 & 0.59 &0.72 &  0.72& 0.67  & 0.67 \\
        Platform & 0.58 & 0.58 & 0.72 & 0.72 & 0.71 & 0.71\\
        Mozilla & 0.50 &0.50 &0.65 &0.65 &0.64&0.64\\
        JDT & 0.59 & 0.59& 0.71 &0.71 & 0.68  & 0.69\\
        Columba & 0.62 & 0.62& 0.73 & 0.73 & 0.79  & 0.79 \\
        PostgreSQL & 0.60 &0.60&0.74&0.74&0.73 &0.73\\ \hline
        Average & 0.58 & 0.58 & 0.71 & 0.71 & 0.70 &0.70 \\ \hline
    \end{tabular}
    % }
    \label{tab:comp_Popt}
\end{table}
\begin{table}[!htp]
    \centering
    \caption{Comparison in $Recall$ : Yang's method~(A) vs. our implementation~(B)}
    % \resizebox{0.48\textwidth}{!}{
    \begin{tabular}{l|c c| c c| c c}
    \hline
        \multirow{2}{*}{Project} &
        \multicolumn{2}{c|}{EALR} &
        \multicolumn{2}{c|}{LT} &\multicolumn{2}{c}{AGE} \\
        \cline{2-7}
        &A & B& A&B&A&B \\
        \hline
        Bugzilla & 0.29 & 0.30 &0.45 &  0.45& 0.38  & 0.38 \\
        Platform & 0.31 & 0.30 & 0.43 & 0.43 & 0.43 & 0.43\\
        Mozilla & 0.18 &0.18 &0.36 &0.36 &0.28&0.28\\
        JDT & 0.32 & 0.34& 0.45 &0.45 & 0.41  & 0.41\\
        Columba & 0.40 & 0.42& 0.44 & 0.44 & 0.57  & 0.57 \\
        PostgreSQL & 0.36 &0.36&0.43&0.43&0.43 &0.43\\ \hline
        Average & 0.31 & 0.32 & 0.43 & 0.43 & 0.41 & 0.41 \\ \hline
    \end{tabular}
    % }
    \label{tab:comp_Recall}
\end{table}
\textbf{RQ1: Do all unsupervised predictors perform better than supervised predictors?}

To answer this question, we build four supervised predictors and twelve
unsupervised predictors on the six project data sets using incremental learning
method as described in Section~\ref{exp_design}. 
% For each project, every 6 consecutive
% months data is used for one experiment, where the first 2 months  data are used train
% the supervised defect learners and the last two 2 months data is used as 
% the test data to evaluate the performance for both supervised and unsupervised predictors.

\begin{figure*}[!htbp]
\begin{center}
    \includegraphics[width=.5\textwidth,height=2.6in]{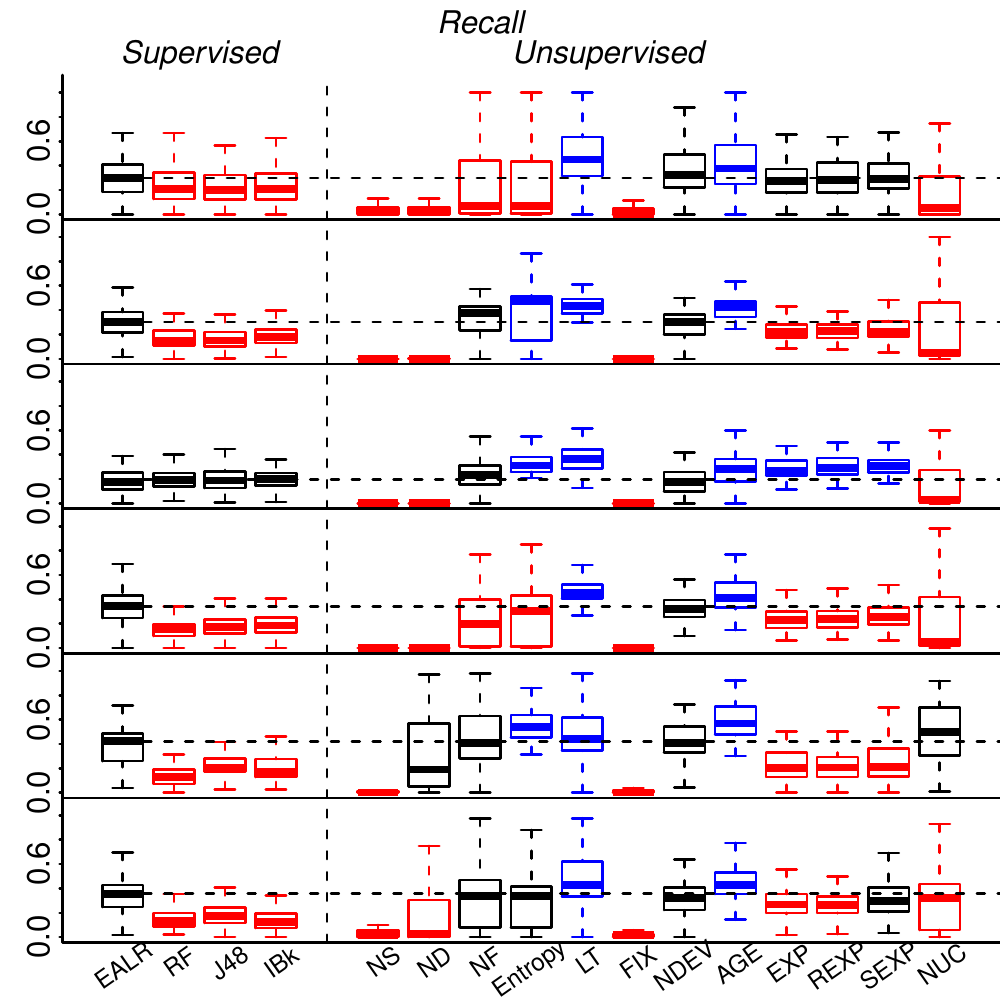}\includegraphics[width=.5\textwidth,height=2.6in]{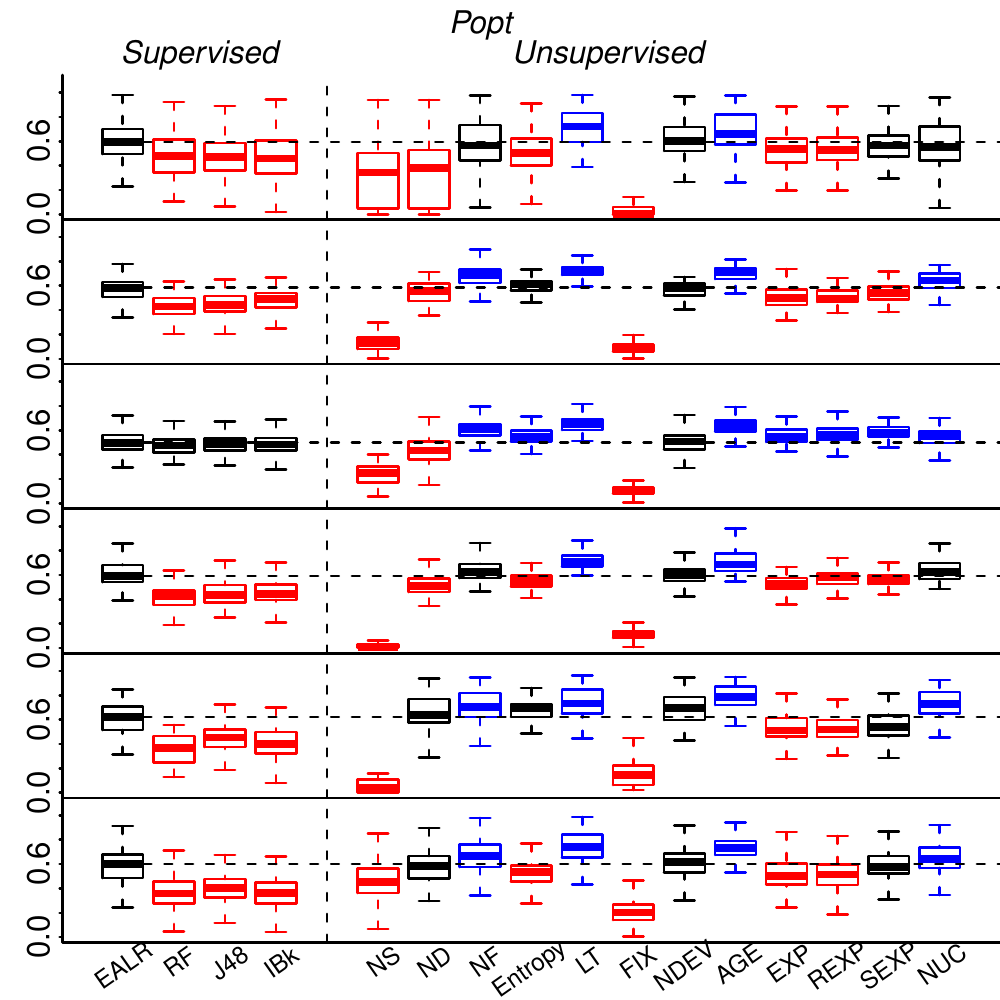}
    \includegraphics[width=.5\textwidth,height=2.6in]{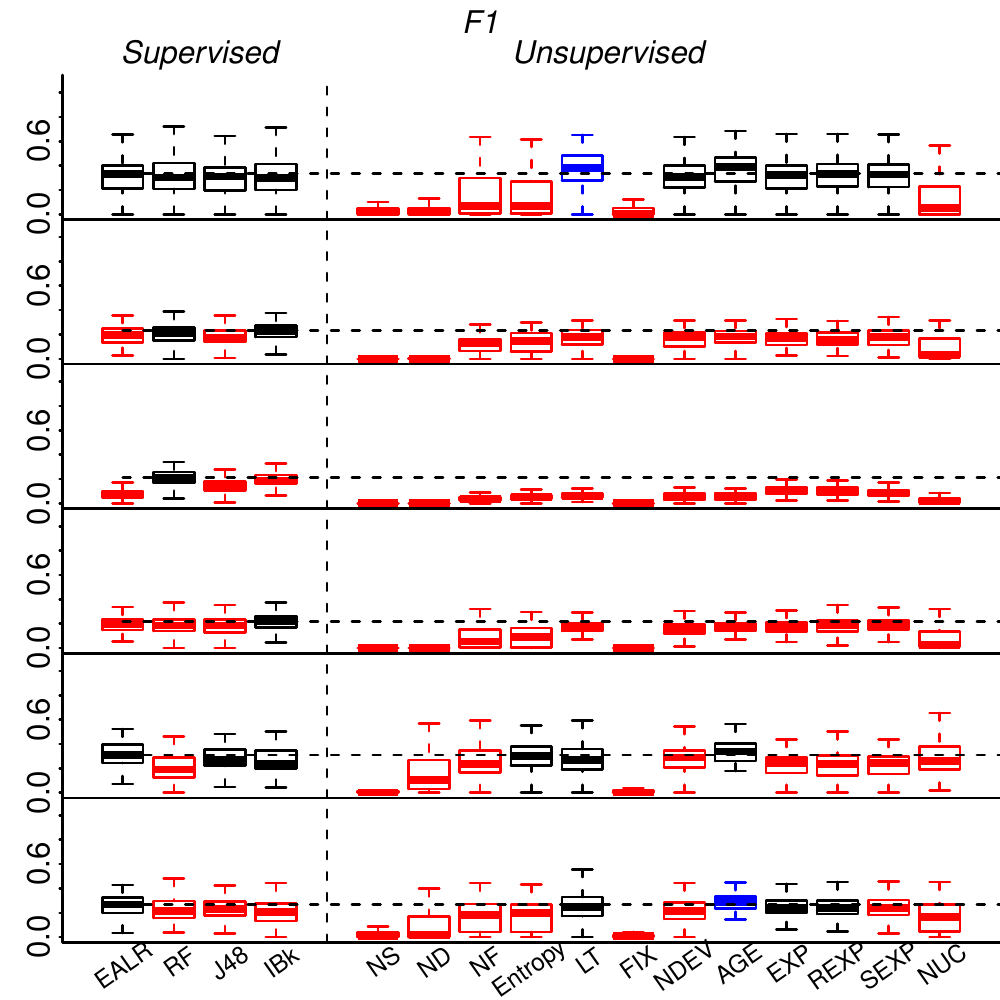}\includegraphics[width=.5\textwidth,height=2.6in]{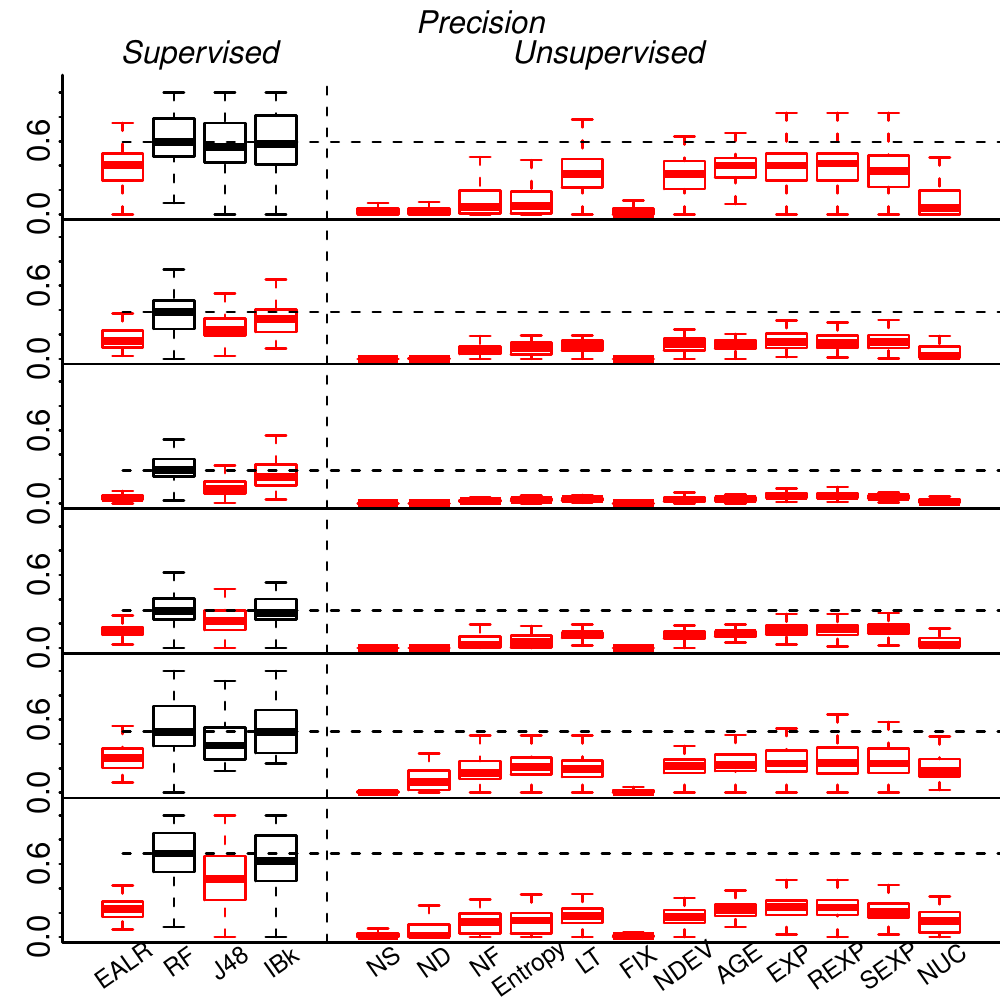}
\caption{Performance comparisons between 
supervised and  unsupervised predictors over six projects~(from top to bottom are  Bugzilla, Platform, Mozilla, JDT, Columba, PostgreSQL). } 
\label{fig:sup_unsup2}
\end{center}
\end{figure*}

 \fig{sup_unsup2} shows the boxplot of {\it Recall}, {\it $P_{opt}$}, {\it F1} and {\it Precision} for
supervised predictors and unsupervised predictors on all data sets. For each predictor, the boxplot shows the
25th percentile, median and 75 percentile  values for one data set. The horizontal dashed lines indicate
the median of the best supervised predictor, which is to help visualize the median differences between unsupervised
predictors and supervised predictors.

% \begin{figure*}[!htbp]
% \begin{center}
%     \includegraphics[width=.5\textwidth,height=2.7in]{pic/hist-cv-tw-Sup-Unsup-ACC.pdf}\includegraphics[width=.5\textwidth,height=2.7in]{pic/hist-cv-tw-Sup-Unsup-Popt.pdf}
% \caption{{\it Recall} and $P_{opt}$ results.
% Comparing  supervised and unsupervised predictors over six projects (from top to bottom are  Bugzilla, Platform, Mozilla, JDT, Columba, PostgreSQL). } 
% \label{fig:sup_unsup1}
% \end{center}
% \end{figure*}

The colors of the boxes within \fig{sup_unsup2} indicate
the significant difference between learners:
\bi
\item
The {\bf  blue color} represents that the corresponding  unsupervised predictor
is significantly {\it better} than the best supervised predictor according to Wilcoxon signed-rank, where the
BH corrected p-value is less than 0.05 {\it and} the magnitude of the difference between these two learners is NOT trivial
according to Cliff's delta, where $|\delta| \geq 0.147$.
\item
The {\bf black color} represents that the corresponding unsupervised predictor is not significantly {\it better} than the best supervised predictor {\it or} the magnitude of the difference between these two learners is trivial, where $|\delta| \leq 0.147$.
\item
The {\bf red color} represents that the corresponding unsupervised predictor is significantly {\it worse} than the best supervised predictor {\it and} the  magnitude of the difference between these two learners is NOT trivial.
\ei
From   \fig{sup_unsup2},  we can clearly see that not all unsupervised
predictors perform statistically better than the best supervised predictor
across all different evaluation metrics. Specifically, for {\it Recall}, 
on one hand, there are only $\frac{2}{12}$,
$\frac{3}{12}$, $\frac{6}{12}$, $\frac{2}{12}$, $\frac{3}{12}$ and $\frac{2}{12}$ of
all unsupervised predictors that perform statistically {\it better} than
the best supervised predictor on six data sets, respectively.
On the other hand, there are $\frac{6}{12}$, $\frac{6}{12}$, $\frac{4}{12}$, $\frac{6}{12}$,
$\frac{5}{12}$ and $\frac{6}{12}$ of
all unsupervised predictors perform statistically {\it worse} than the best supervised predictor on the six data sets, respectively. This indicates that:
\bi
\item
About $50\%$ of the unsupervised predictors   
perform worse than the best supervised predictor on any data set;
\item
Without any prior knowledge,  we can not  know which
unsupervised predictor(s)   works adequately on the testing data. 
\ei
Note that the above two
points from   {\it Recall} also hold for $ P_{opt}$.

For {\it F1}, we see that only LT on Bugzilla and AGE on PostgreSQL
 perform statistically better than the best supervised predictor. 
 Other than that, no unsupervised predictor performs better on any data set. Furthermore, surprisingly,
 no unsupervised predictor works significantly better than the best supervised predictor on any 
data sets in terms of {\it Precision}. As we can see, Random Forests performs well on all six data sets.
This suggests that unsupervised predictors have very low
precision for effort-aware defect prediction and can not be deployed to any business situation where precision is critical.

\begin{figure*}[htbp]
\begin{center}
 \includegraphics[width=.5\textwidth,height=2.6in]{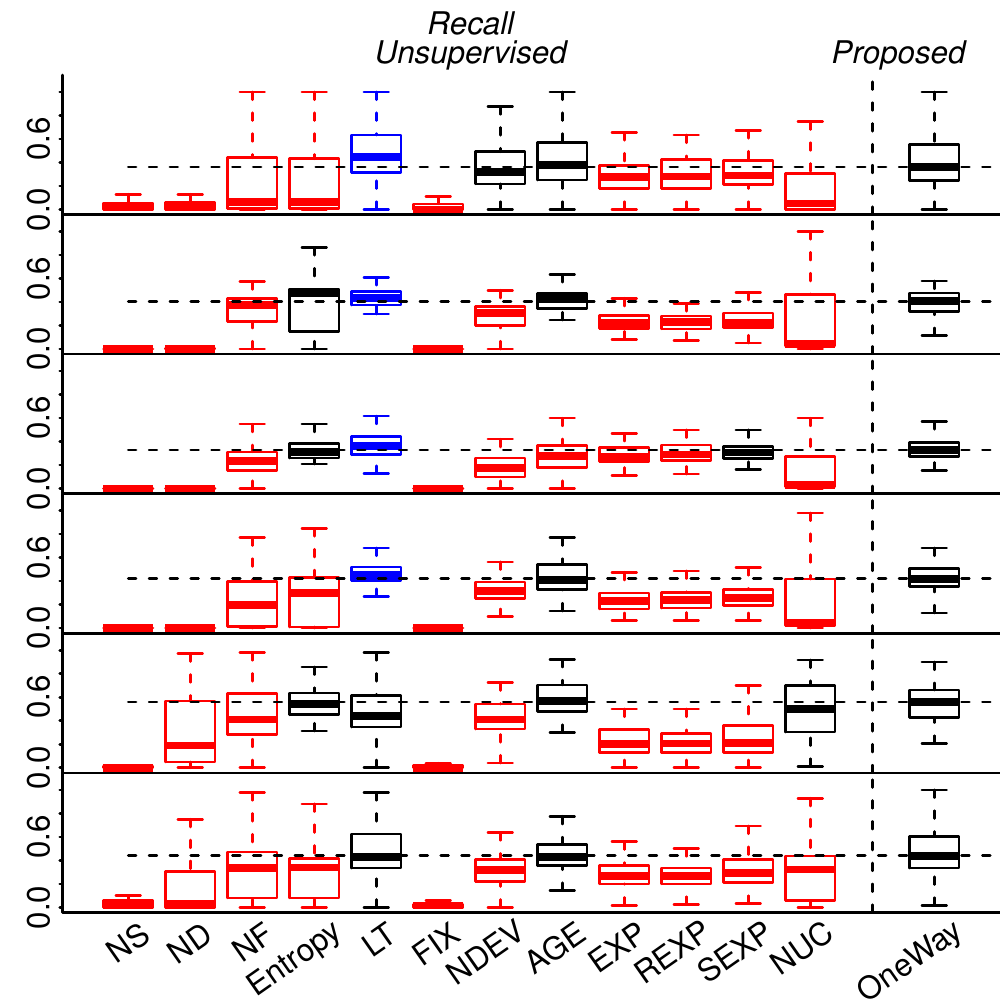}\includegraphics[width=.5\textwidth,height=2.6in]{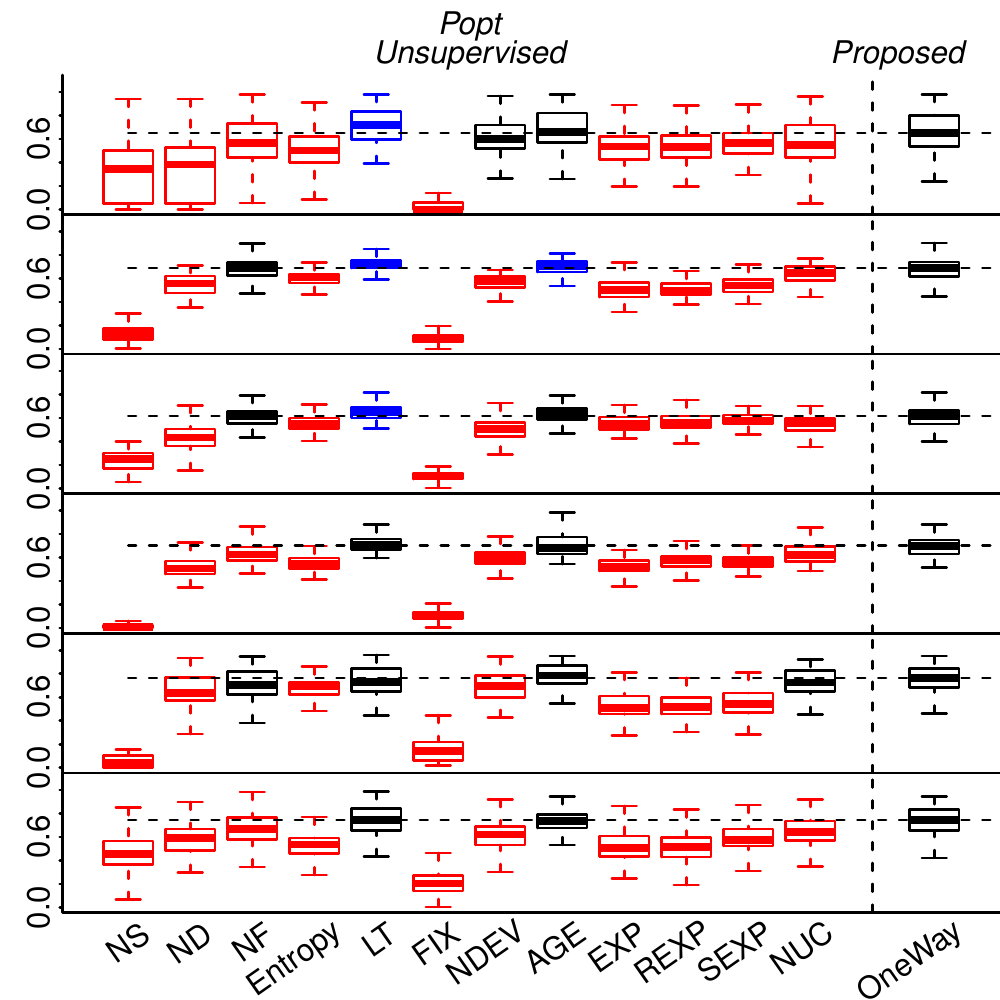}
 \includegraphics[width=.5\textwidth,height=2.6in]{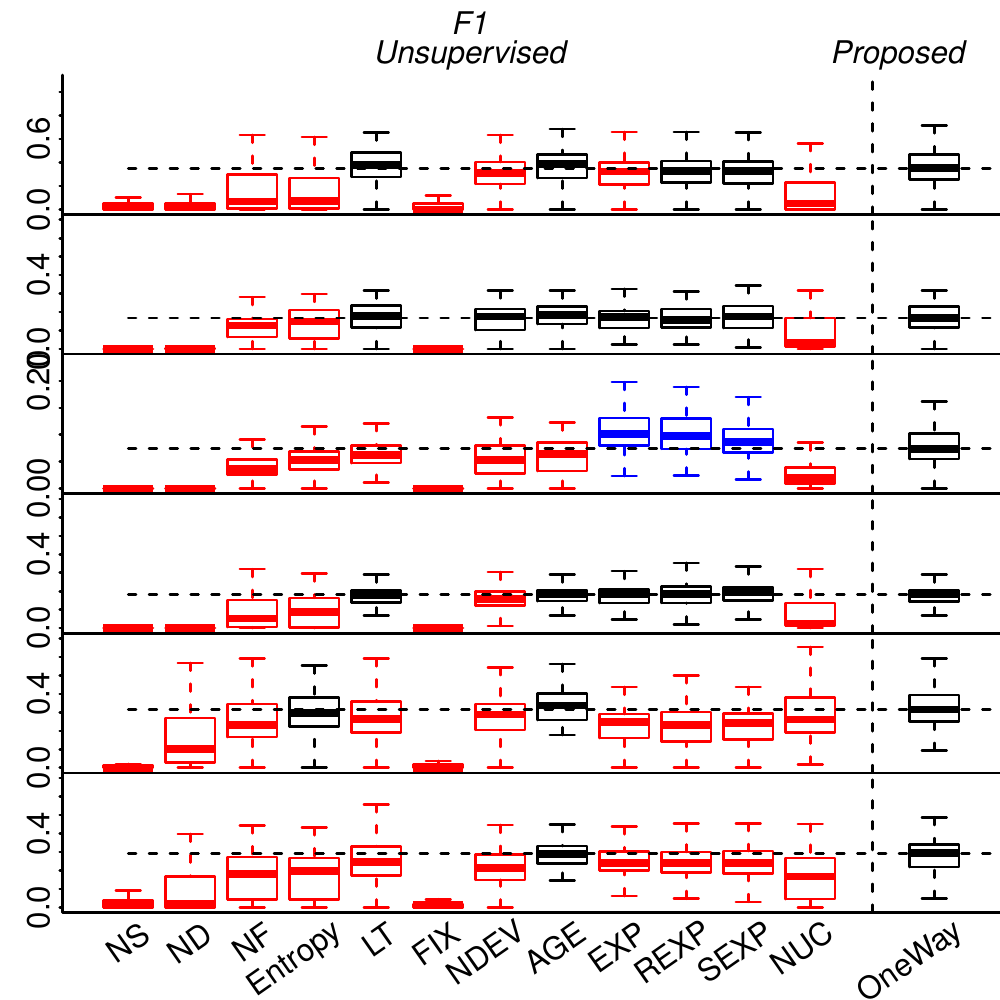}\includegraphics[width=.5\textwidth,height=2.6in]{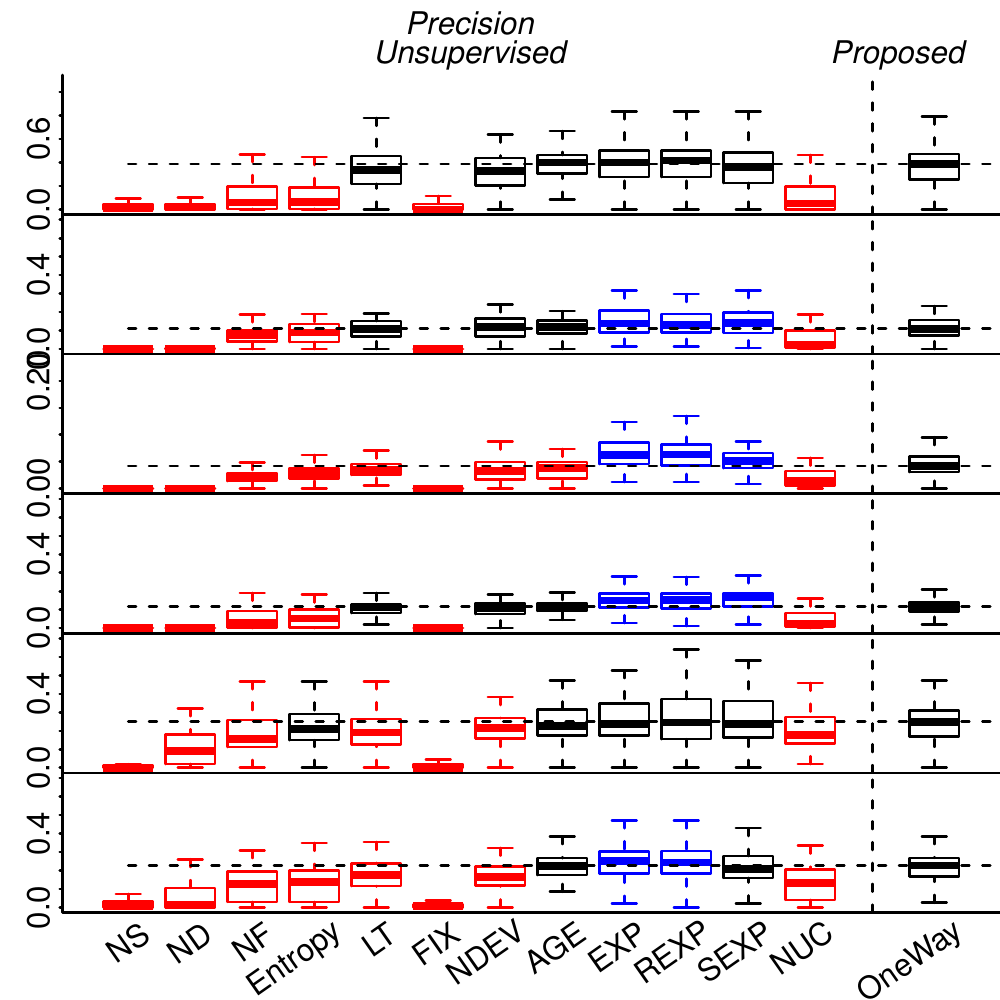}
\caption{Performance comparisons between the proposed {\it OneWay} learner and unsupervised predictors over six projects~(from top to bottom are Bugzilla, Platform, Mozilla, JDT, Columba, PostgreSQL).}
\label{fig:unsup_oneway}
\end{center}
\end{figure*}

Overall, for a given data set, no one specific unsupervised predictor works 
better than the best supervised predictor across all evaluation
metrics. For a given measure, most unsupervised predictors
did not perform better across all data sets. 
In summary: 
\vskip 1ex
 \begin{myshadowbox}
    Not all unsupervised predictors perform better than supervised predictors for each project and for different evaluation measures.
 \end{myshadowbox}
 
Note the implications of this finding: some extra knowledge is required to prune the worse unsupervised models, such as the knowledge that can come from labelled data.
Hence, we must conclude the opposite to Yang et al.; i.e. some supervised labelled data 
must be applied before we can reliably deploy unsupervised defect 
predictors on testing data. 

\textbf{RQ2: Is it beneficial to use supervised data to prune away all but one of the Yang et al. predictors?}

To answer this question, we compare the {\it OneWay} learner with all twelve
unsupervised predictors. All these predictors are tested on  the six project data 
sets using the same experiment scheme as we did in RQ1.

\begin{figure*}[htbp]
\begin{center}
 \includegraphics[width=.25\textwidth,height=2.6in]{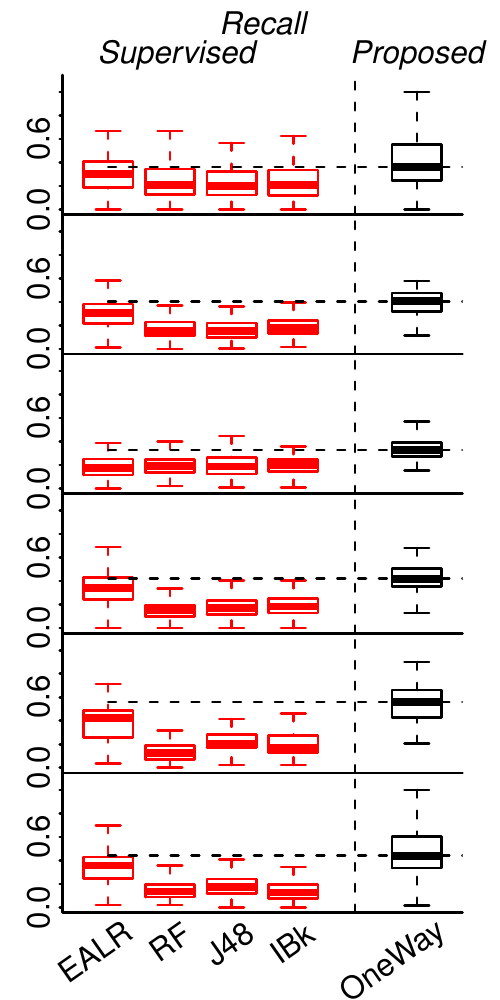}\includegraphics[width=.25\textwidth,height=2.6in]{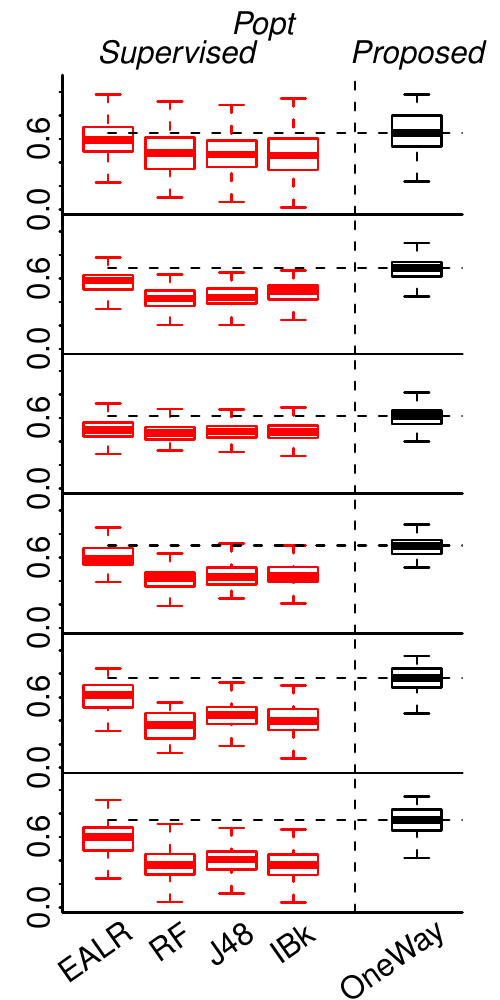}\includegraphics[width=.25\textwidth,height=2.6in]{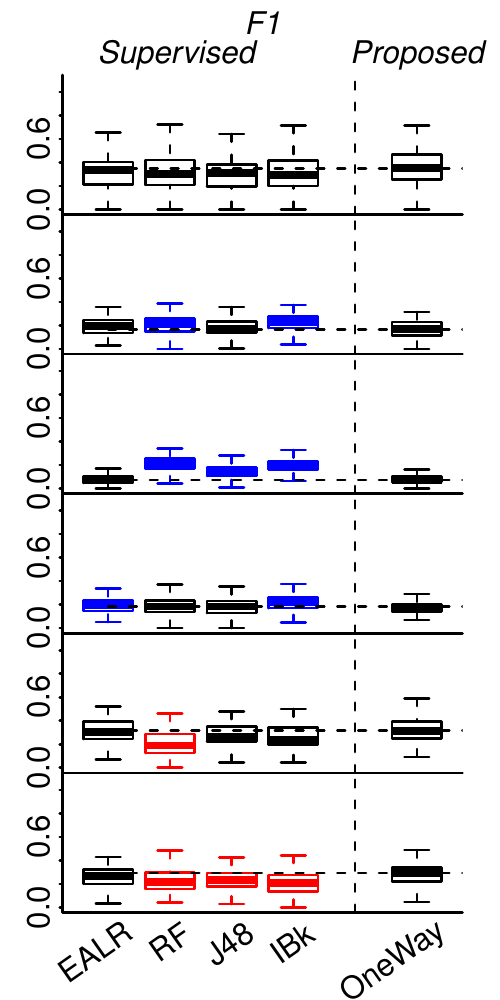}\includegraphics[width=.25\textwidth,height=2.6in]{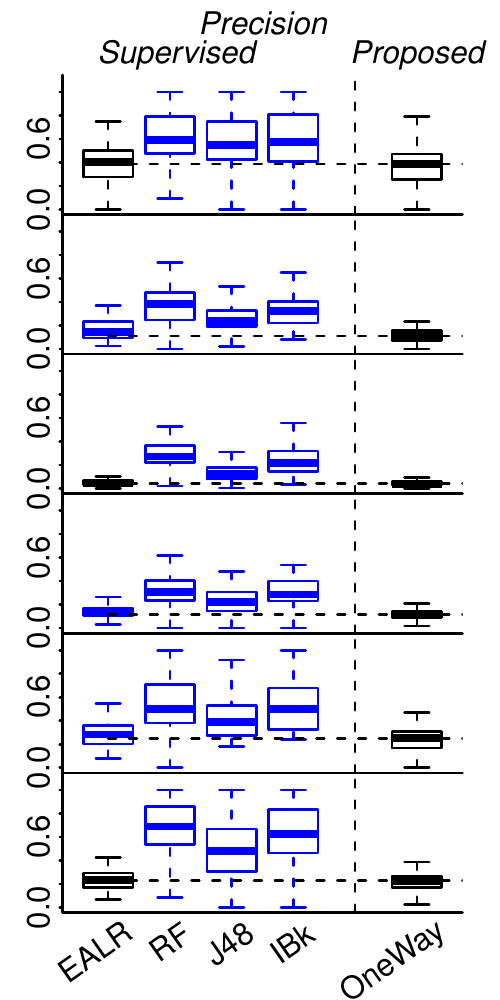}
\caption{Performance comparisons between the proposed {\it OneWay} learner and supervised predictors over six projects~(from top to bottom are  Bugzilla, Platform, Mozilla, JDT, Columba, PostgreSQL).}
\label{fig:sup_oneway}
\end{center}
\end{figure*}

\fig{unsup_oneway} shows the boxplot for the performance distribution of
unsupervised predictors and the proposed
{\it OneWay} learner on six data sets across four evaluation measures. The horizontal 
dashed line denotes the median value of {\it OneWay}. Note that in \fig{sup_oneway}, blue 
means this learner is statistically better than {\it OneWay}, red means worse, 
and black means no difference. As we can see, in   {\it Recall},
only one unsupervised predictor, LT, outperforms {\it OneWay} in $\frac{4}{6}$ data sets. However, {\it OneWay}
significantly outperform$\frac{9}{12}$, $\frac{9}{12}$, $\frac{9}{12}$, $\frac{10}{12}$,
 $\frac{8}{12}$ and $\frac{10}{12}$ of total unsupervised predictors on six data sets, respectively.
 This observation indicates that {\it OneWay} works significantly better 
 than almost all learners on all 6 data sets in terms of {\it Recall}.
 
 Similarly, we observe that only LT predictor works better than {\it OneWay} in
$\frac{3}{6}$ data sets in terms of {\it $P_{opt}$} and {\it AGE} outperforms {\it OneWay} only on the {\it platform} data set.
For the remaining experiments, {\it OneWay} performs better than all the other predictors~(on average, 9 out of 12 predictors). 

In addition, according to {\it F1}, only three unsupervised predictors EXP/REXP/SEXP 
perform better than {\it OneWay} on the Mozilla data set and
LT predictor just performs as well as {\it OneWay} 
~(and has no  advantage over {\it OneWay}). We note that similar findings can be observed in {\it Precision} measure.

\tab{unsu_oneway} provides the median values of the best unsupervised predictor compared with {\it OneWay}
for each evaluation measure on all data sets. Note that, in practice, we can not know which unsupervised predictor is the best out of the 12 unsupervised predictors by Yang et al.'s method before we access to the labels of testing data. In other words, to aid our analysis, the best unsupervised ones  in \tab{unsu_oneway} are selected when referring to the true labels of testing data, which are not available in practice. In that table, for each evaluation measure,
the number in {green} cell indicates that the best unsupervised predictor has a large advantage
over {\it OneWay} according to the Cliff's $\delta$; Similarly, the {yellow} cell means medium advantage
and the {gray} cell means small advantage. 

From \tab{unsu_oneway}, we observe that out of 24 experiments on all evaluation measures, 
{\it none} of these best unsupervised predictors outperform {\it OneWay} with a large advantage according
to the Cliff's $\delta$. Specifically,  according to {\it Recall} and $P_{opt}$,
even though the best unsupervised predictor, LT, outperforms {\it OneWay} on four and three data sets, all of
these advantage are small. Meanwhile, REXP and EXP have a medium improvement
over {\it OneWay} on one and two data sets for {\it F1} and {\it Precision}, respectively.
In terms of the average scores, the maximum magnitude of the difference between the best unsupervised 
learner and {\it OneWay} is $0.02$. In other words, {\it OneWay} is comparable with the best unsupervised predictors on all data sets for all evaluation measures even though the best unsupervised predictors might not be known before testing.

\begin{table}[]
    \centering
    \caption{Best unsupervised predictor~(A) vs. OneWay~(B). The colorful cell indicates the size effect: green for large; yellow for medium; gray for small.}
    \resizebox{0.48\textwidth}{!}{
    \begin{tabular}{l|c c| c c| c c| c c}
    \hline
        \multirow{2}{*}{Project} &
        \multicolumn{2}{c|}{{\it Recall}} &
        \multicolumn{2}{c|}{$P_{opt}$} &
        \multicolumn{2}{c|}{{\it F1}} &
        \multicolumn{2}{c}{{\it Precision}}\\
        \cline{2-9}
        &A~(LT) & B & A~(LT)&B&A~(REXP)&B &A~(EXP)&B \\
        \hline
        Bugzilla &\colorbox{gray!80}{0.45} & 0.36 &\colorbox{gray!80}{0.72} &  0.65& 0.33  & 0.35 & 0.40 & 0.39  \\
        Platform & \colorbox{gray!80}{0.43} & 0.41 & \colorbox{gray!80}{0.72} & 0.69 & 0.16 & 0.17 &\colorbox{gray!80}{0.14} &0.11\\
        Mozilla & \colorbox{gray!80}{0.36} &0.33 &\colorbox{gray!80}{0.65} &0.62 &\colorbox{yellow}{0.10}&0.08 &\colorbox{yellow}{0.06} &0.04\\
        JDT & \colorbox{gray!80}{0.45}& 0.42& 0.71 & 0.70  & 0.18 & 0.18 &\colorbox{yellow}{0.15} &0.12\\
        Columba & 0.44& 0.56 & 0.73 & 0.76 & 0.23 & 0.32 &0.24 &0.25\\
        PostgreSQL & 0.43 &0.44&0.74&0.74&0.24 & 0.29 &\colorbox{gray!80}{0.25} &0.23\\ \hline
        Average & 0.43 &0.42 &0.71& 0.69 & 0.21 & 0.23 &0.21 &0.19\\ \hline
    \end{tabular}
    }
    \label{tab:unsu_oneway}
\end{table}

Overall, we find that (1) no one unsupervised predictor significantly
outperforms {\it OneWay} on all data sets for a given evaluation measure; (2) mostly, {\it OneWay}
works as well as the best unsupervised predictor and has significant better performance than almost
all unsupervised predictors on all data sets for all evaluation measures. Therefore, the above results suggest:
\vskip 1ex
 \begin{myshadowbox}
    As a simple supervised predictor, {\it OneWay} has competitive performance  and
    it performs  better than most unsupervised predictors for effort-aware JIT defect prediction. 
 \end{myshadowbox}

Note the implications of this finding: the supervised learning
utilized in {\it OneWay} can significantly outperform the
unsupervised models.

\begin{table}[]
    \centering
    \caption{Best supervised predictor~(A) vs. OneWay~(B). The colorful cell indicates the size effect: green for large; yellow for medium; gray for small.}
    \resizebox{0.48\textwidth}{!}{
    \begin{tabular}{l|c c| c c| c c| c c}
    \hline
        \multirow{2}{*}{Project} &
        \multicolumn{2}{c|}{{\it Recall}} &
        \multicolumn{2}{c|}{$P_{opt}$} &
        \multicolumn{2}{c|}{{\it F1}} &
        \multicolumn{2}{c}{{\it Precision}}\\
        \cline{2-9}
        &A~(EALR) & B & A~(EALR)&B&A~(IBk)&B &A~(RF)&B \\
        \hline
        Bugzilla & 0.30 & 0.36 &0.59 &  0.65& 0.30  & 0.35 & \colorbox{green!55!blue}{0.59} & 0.39  \\
        Platform & 0.30 & 0.41 & 0.58 & 0.69 & \colorbox{yellow}{0.23} & 0.17 &\colorbox{green!55!blue}{0.38} &0.11\\
        Mozilla & 0.18 &0.33 &0.50 &0.62 &\colorbox{green!55!blue}{0.18}&0.08 &\colorbox{green!55!blue}{0.27} &0.04\\
        JDT & 0.34& 0.42& 0.59 & 0.70  & \colorbox{gray!80}{0.22} & 0.18 &\colorbox{green!55!blue}{0.31} &0.12\\
        Columba & 0.42& 0.56 & 0.62 & 0.76 & 0.24 & 0.32 &\colorbox{green!55!blue}{0.50} &0.25\\
        PostgreSQL & 0.36 &0.44&0.60&0.74&0.21 & 0.29 &\colorbox{green!55!blue}{0.69} &0.23\\ \hline
        Average & 0.32 &0.42 &0.58& 0.69 & 0.23 & 0.23 &0.46 &0.19\\ \hline
    \end{tabular}
    }
    \label{tab:unsup_oneway}
\end{table}

\textbf{RQ3: Does {\it OneWay} perform  better than more complex standard supervised predictors?}

To answer this question, we compare {\it OneWay} learner with four
supervised predictors, including EALR, Random Forests, J48 and IBk. 
EALR is considered to be state-of-the-art learner for effort-aware
JIT defect prediction~\cite{yang2016effort,kamei2013large} and all the other three learners are widely used 
in defect prediction literature over past years~\cite{lessmann2008benchmarking,fu2016tuning,kamei2013large,fukushima2014empirical,turhan2009relative,hall2012systematic}. We evaluate all these
learners on the six project data 
sets using the same experiment scheme as we did in RQ1.

From \fig{sup_oneway}, we have the following observations. Firstly, 
the performance of {\it OneWay} is significantly better than all
these four supervised predictors in terms of {\it Recall} and ${P_{opt}}$
on all six data sets. Also, EALR works better than Random Forests, J48 and IBk,
which is consistent with Kamei et al's finding~\cite{kamei2013large}. 

Secondly,
according to {\it F1}, Random Forests and IBk perform slightly better than {\it OneWay}
in two out of six data sets. For most cases, {\it OneWay} has a similar performance
to these supervised predictors and there is not much difference between them.

However,
when reading {\it Precision} scores, we find that, in most cases, supervised
learners perform significantly better than {\it OneWay}. Specifically, Random Forests,
J48 and IBk outperform {\it OneWay} on all data sets and EALR is better on three data sets. 
This finding is consistent with the observation in RQ1 where all unsupervised predictors
perform worse than supervised predictors for {\it Precision}. 
% As an aside, EALR is never
% the best supervised predictor in any case, but it is still better or simiar to {\it OneWay}.

From \tab{unsup_oneway},  we have the following observation.
First of all, in terms of {\it Recall} and ${P_{opt}}$, the maximum  difference in median values between 
EALR and {\it OneWay} are $0.15$ and $0.14$, respectively, which are $83\%$ and $23\%$ improvements over $0.18$ and $0.60$ on  Mozilla and PostgreSQL data sets.
For both measures, {\it OneWay} improves the average scores by $0.1$ and $0.11$, which are
$31\%$ and $19\%$ improvement over EALR. Secondly, according to {\it F1}, IBk outperforms
{\it OneWay}  on three data sets with a large, medium and small advantage, respectively.
The largest difference in median is $0.1$. Finally, as we discussed before, the best 
supervised predictor for {\it Precision}, Random Forests, has a very large advantage over
{\it OneWay} on all data sets. The largest difference is $0.46$ on PostgreSQL data set.

Overall, according to the above analysis, we conclude that:
\vskip 1ex
 \begin{myshadowbox}
    % {\it OneWay} performs significantly better than all four supervised predictors in terms of {\it Recall}
    % and $P_{opt}$, competitive for {\it F1}, but not for {\it Precision}.
    \textit{OneWay} performs significantly better than all four supervised
    learners in terms of {\it Recall} and $P_{opt}$; It performs just as well as other learners for {\it F1}. As for Precision, other supervised predictors outperform \textit{OneWay}.
 \end{myshadowbox}

Note the implications of this finding: simple tools like {\em OneWay}
perform adequately but for all-around performance, more sophisticated learners are recommended.

As to when to use {\it OneWay} or  supervised predictors like Random Forests, that is  an open question. According to ``No Free Lunch Theorems''\cite{wolpert2002supervised}, no method is always best and we show unsupervised predictors are often worse on a project-by-project basis. So ``best'' predictor selection is  a matter of local assessment, requiring labelled training data (an issue ignored by Yang et al).

% \section{Discussion}
% \subsection{ Is a Low Precision {\it OneWay} Learner Useful}
% \wei{DO we need to use precision? how to explain. feel struggled here.}

% In last section, we notice that {\it OneWay} learner does not 
% perform well in terms of {\it Precision} compared to the supervised predictors
% in RQ3. However, it is still a quite useful learner and the reason is as follows.

% First of all, for effort-ware defect prediction like this study, we only set 
%  $20\%$ efforts(measured in code churn) as our budget, which is quite small amount
%  of effort compared with the total. 

\section{Threats to Validity}\label{threats}
 
\textbf{Internal Validity}. The internal validity is related to uncontrolled
aspects that may affect the experimental results. One threat to the internal validity
is how well our implementation of unsupervised predictors could represent 
the Yang et al.'s method. To mitigate this threat, based on Yang et al ``R''code, we strictly follow 
the approach described in Yang et al's work and test our implementation on the
same data sets as in Yang et al.~\cite{yang2016effort}. By comparing the performance
scores, we find that our implementation can generate the same results. Therefore,
we believe we can avoid this threat.

\textbf{External Validity}. The external validity is related to the
possibility to generalize our results. Our observations and conclusions
from this study may not be generalized to other software projects. In this study,
we use six widely used open source software project data as the subject.
As all these software projects are written in java, we can not  guarantee that
our findings can be directly generalized to other projects,
specifically to the software that implemented in other programming
languages. Therefore, the future work might include to verify our findings on 
other software project.

In this work, we used the data sets from~\cite{yang2016effort,kamei2013large}, where
totally 14 change metrics were extracted from the software projects.  We build and test the
{\it OneWay} learner on those metrics as well. However, there might be some other metrics
that not measured in these data sets that work well as indicators for defect prediction.
For example, when the change was committed (e.g., morning, afternoon or evening),
functionality of the the files modified in this change~(e.g., core functionality or not).
Those new metrics that are not explored in this study might improve the performance of 
our {\it OneWay} learner.

\section{Conclusion and Future Work}\label{conclusion}

This paper replicated and refutes Yang et al.'s results~\cite{yang2016effort}
on unsupervised predictors for effort-ware just-in-time 
defect prediction. Not all unsupervised predictors
work better than supervised predictors (on all six data sets, for different evaluation
measures). This suggests that we can not randomly pick an  unsupervised predictor
to perform effort-ware JIT defect prediction.  
Rather, it is necessary to use supervised methods to  pick   best models before deploying them to a project.
For that task, supervised predictors like {\it OneWay} are useful
to automatically
select the potential best model.

In the above, {\em OneWay} peformed very well for   {\it Recall}, $P_{opt}$ and {\it F1}. Hence, it must be asked: ``Is defect
prediction inherently simple? And does it need
anything other than {\em OneWay}?''. In this context,
it is useful to recall that  {\em OneWay}'s results
for precision were not competitive.
Hence we say, that if learners are to be deployed in domains
where precision is critical, then {\em OneWay} is too simple. 

This study opens the new research direction of applying simple supervised 
techniques to perform defect prediction. As shown in this
study as well as Yang et al.'s work~\cite{yang2016effort}, instead of using
traditional machine learning algorithms like J48 and Random Forests, simply sorting
data according to one metric can be a good defect predictor model, at least for
effort-aware just-in-time defect prediction. Therefore, we recommend the future
defect prediction research should focus more on simple techniques.

For the future work, we plan to extend this study on other software projects,
especially those developed by the other programming languages. After that,
we plan to investigate new change metrics to see if that helps improve
{\it OneWay}'s performance.

\section{Addendum}
As this paper was going to press, we learned of  new
 papers that updated  the Yang et al. study: 
Liu et al.
at EMSE'17~\cite{liu2017code} and Huang et al. at 
ICSMSE'17~\cite{huang17}. 
We thank these authors
for the courtesy of sharing a
pre-print of those new results. We also thank them for 
using 
concepts from a  pre-print of our paper in their work\footnote{
Our pre-print was posted to arxiv.org March 1, 2017. We hope more
researchers will use tools like arxiv.org to speed along the pace of software research.}.
Regretfully, we have yet to return those favors:
due to deadline pressure, we have not 
been able to confirm   their results. 

As to  technical specifics, Liu et al.  use a single {\em churn}   measure
(sum of number of lines added and deleted)
to build an unsupervised predictors that  does remarkably better than 
{\em OneWay} and EARL (where the latter  could access
all the variables).  
While this result is currently unconfirmed, 
it could well have ``raised the bar''
for unsupervised defect prediction. Clearly,
 more experiments are needed in this area.  
For example, when comparing the Liu et al. methods
to  {\em OneWay} and standard supervised learners, we could (a)~give
all learners access to the churn variable;  (b)~apply the Yang   transform of $\frac{1}{M(c)}$ to all variables prior to learning;
(c)~use  more elaborate supervised methods including  synthetic minority over-sampling~\cite{Agrawa17} and automatic hyper-parameter optimization~\cite{fu2016tuning}.

\section*{ACKNOWLEDGEMENTS}
The work is partially funded by an NSF award \#1302169.

\balance
\bibliographystyle{plain}
\bibliography{unsuper.bbl}

\end{document}